\documentclass[%
 reprint,
superscriptaddress,
showpacs,preprintnumbers,
 amsmath,amssymb,
 aps,
prl,
floatfix,
]{revtex4-1}

\usepackage{graphicx}
\usepackage{dcolumn}
\usepackage{bm}
\usepackage{xcolor}

\usepackage{hyperref}

\begin{document}
\title{Distribution of Squeezed States through an Atmospheric Channel}
\author{Christian Peuntinger}\thanks{contributed equally to this work\\correspondance to: christian.peuntinger@mpl.mpg.de, \\or christoph.marquardt@mpl.mpg.de}
\affiliation{Max Planck Institute for the Science of Light, G\"nther-Scharowsky-Stra{\ss}e 1/Building 24, 91058 Erlangen, Germany}
\affiliation{Institute of Optics, Information and Photonics, University of Erlangen-Nuremberg (FAU), Staudtstra{\ss}e 7/B2, 91058 Erlangen, Germany}
\author{Bettina Heim}\thanks{contributed equally to this work\\correspondance to: christian.peuntinger@mpl.mpg.de, \\or christoph.marquardt@mpl.mpg.de}
\affiliation{Max Planck Institute for the Science of Light, G\"nther-Scharowsky-Stra{\ss}e 1/Building 24, 91058 Erlangen, Germany}
\affiliation{Institute of Optics, Information and Photonics, University of Erlangen-Nuremberg (FAU), Staudtstra{\ss}e 7/B2, 91058 Erlangen, Germany}
\affiliation{Erlangen Graduate School in Advanced Optical Technologies (SAOT), FAU, Paul-Gordan-Stra{\ss}e 6, 91052 Erlangen, Germany}
\author{Christian R. M\"uller}
\affiliation{Max Planck Institute for the Science of Light, G\"nther-Scharowsky-Stra{\ss}e 1/Building 24, 91058 Erlangen, Germany}
\affiliation{Institute of Optics, Information and Photonics, University of Erlangen-Nuremberg (FAU), Staudtstra{\ss}e 7/B2, 91058 Erlangen, Germany}
\author{Christian Gabriel}
\affiliation{Max Planck Institute for the Science of Light, G\"nther-Scharowsky-Stra{\ss}e 1/Building 24, 91058 Erlangen, Germany}
\affiliation{Institute of Optics, Information and Photonics, University of Erlangen-Nuremberg (FAU), Staudtstra{\ss}e 7/B2, 91058 Erlangen, Germany}
\author{Christoph Marquardt}
\affiliation{Max Planck Institute for the Science of Light, G\"nther-Scharowsky-Stra{\ss}e 1/Building 24, 91058 Erlangen, Germany}
\affiliation{Institute of Optics, Information and Photonics, University of Erlangen-Nuremberg (FAU), Staudtstra{\ss}e 7/B2, 91058 Erlangen, Germany}
\affiliation{Erlangen Graduate School in Advanced Optical Technologies (SAOT), FAU, Paul-Gordan-Stra{\ss}e 6, 91052 Erlangen, Germany}\author{Gerd Leuchs}
\affiliation{Max Planck Institute for the Science of Light, G\"nther-Scharowsky-Stra{\ss}e 1/Building 24, 91058 Erlangen, Germany}
\affiliation{Institute of Optics, Information and Photonics, University of Erlangen-Nuremberg (FAU), Staudtstra{\ss}e 7/B2, 91058 Erlangen, Germany}
\affiliation{Erlangen Graduate School in Advanced Optical Technologies (SAOT), FAU, Paul-Gordan-Stra{\ss}e 6, 91052 Erlangen, Germany}

\date{\today}

\begin{abstract}
Continuous variable quantum states of light are used in quantum information protocols and quantum metrology and known to degrade with loss and added noise. We were able to show the distribution of bright polarization squeezed quantum states of light through an urban free-space channel of 1.6\,km length. To measure the squeezed states in this extreme environment, we utilize polarization encoding and a postselection protocol that is taking into account classical side information stemming from the distribution of transmission values. The successful distribution of continuous variable squeezed states is accentuated by a quantum state tomography, allowing for determining the purity of the state.
\end{abstract}
\maketitle

The survival of quantum states in hostile environments is of crucial importance in quantum information processing and of general interest in the understanding of decoherence processes. The transmission of quantum states of light through an atmospheric link is such a demanding task. Free-space quantum state distribution has the potential to form a key component in future quantum networks. It offers the possibility of distributing quantum states with high flexibility in the required infrastructure. Furthermore, links to moving objects, like vehicles, planes, and satellites are conceivable. Recent experiments demonstrated free-space quantum channels in long range ground-to-ground links~\cite{Buttler00,Hughes02,Ursin07,Schmitt-Manderbach07,Fedrizzi09,Yin12} as well as in a link from a ground station to a plane~\cite{Nauerth13}. Other experiments study the feasibility of quantum communication between Earth and space, in general~\cite{Rarity02,Aspelmeyer03,Villoresi08,Armengol08, Bonato09,Meyer-Scott11, Bourgoin13}, or the effects of atmospheric turbulence on quantum states, in particular, and how to exploit them~\cite{Semenov09,Erven12,Capraro12}.
All these experiments (for a general review on free-space quantum communication experiments see Ref.~\cite{Tunick10}), are based on discrete variables, i.e., rely on photon counting at the receiver. On the other hand, recent investigations have proven the feasibility of continuous variable (CV) free-space quantum key distribution~\cite{Elser09,Heim10,Usenko12}. By using homodyne receivers, this approach can offer high speed detection and immunity to stray light. A comprehensive review on CV Gaussian quantum information can be found in Ref.~\cite{Weedbrook12}. Squeezed states can be used in CV quantum key distribution protocols~\cite{Hillery00,Lorenz01} and it has been proven~\cite{Usenko11} that they can provide an enhancement compared to coherent states. When transmitting CV quantum states much care has to be taken to avoid their decoherence. Losses will result in a degradation of the quantum states, whereas in discrete variable systems, these will merely lower the detection rate. Also, phase relations and wave front distortions play an important role, as the applied homodyne detection relies on optical interference and a high homodyne efficiency is crucial. Thus, the safe transmission of CV quantum states in harsh environments poses a major challenge in different physical systems, ranging from optical free-space links to superconducting cavity quantum electrodynamics~\cite{Eichler11}.

In this Letter, we demonstrate the distribution of bright CV squeezed states of light through a turbulent atmospheric channel. The atmosphere introduces fluctuations in amplitude and phase. Because of the fluctuating phase and the resulting wave front distortions, standard homodyne measurements at the receiver are challenging. We circumvent this problem by using polarization encoding. Polarization offers a distinct advantage in this case. The different polarization components can be measured separately, and yet they occupy the same spatial mode and experience identical distortions. Measuring the polarization components averaged over the beam cross section is like a differential measurement, where the distortions cancel out. In that sense, the polarization measurement is analogous to measuring in a decoherence free subspace~\cite{Lidar03, Leuchs09} and is immune to phase fluctuations. Bright polarization squeezed states are equivalent to combining a bright coherent polarization mode and an orthogonally polarized vacuum squeezed mode~\cite{Korolkova02,Heersink06}. Thus, we are able to perform polarization measurements, which can be approximated as homodyne measurements with a bright local oscillator and a polarization multiplexed squeezed vacuum signal. In principle, it would be possible to use a different degree of freedom for the multiplexing of the squeezed signal with the local oscillator or even to use an independent laser at the receiver as a local oscillator. However, it would be technically challenging to achieve a high interferometric visibility. CV polarization encoding also perfectly fits atmospheric transmission systems~\cite{Elser09,Heim10}, and the negligible impact of the atmosphere on the state of polarization was even investigated in a space to ground scenario~\cite{Toyoshima09}. Minor effects occur for the case of ice crystals in the atmosphere, which was not relevant for the urban link measurements reported in this Letter. Additionally, the polarization of the light field is easy to manipulate without introducing considerable loss. Homodyne receivers are, in general, immune to stray light, allowing for daylight operation without any further spectral or spatial filtering.

The atmospheric channel introduces fluctuations of the intensity through beam wandering, distortion, and spreading. Special care is necessary to take into account the resulting variation of the shot noise level during our measurements. We developed a postselection protocol that is also able to enhance the degree of squeezing compared to that of the raw channel data. These results are substantiated by a quantum state reconstruction via a maximum likelihood algorithm~\cite{Vardi93,Hradil97,Lvovsky04,Mueller12}.

\begin{figure}
\centering
\includegraphics[width=0.80\columnwidth]{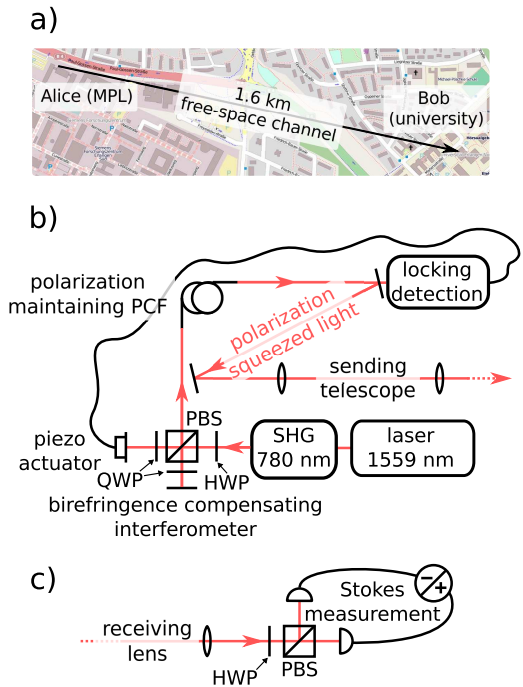}
\caption{Schematic drawing of the experimental setup.
Used abbreviations: SHG, second harmonic generation; HWP, half-wave plate; QWP, quarter-wave plate; PBS, polarizing beam splitter; PCF, photonic crystal fiber. (a) A street map of our free-space channel. The sender {\it (Alice)}, located at the roof of the MPL institute's building, is connected by a 1.6\,km atmospheric transmission line with the receiving station {\it (Bob)}, which is placed in a tall university building. (b) The source of the bright polarization squeezed quantum states. (c) At the receiver, we directly focus onto two detectors after splitting the beam by a combination of a HWP and a PBS. The difference signal of these detectors corresponds to a Stokes observable within the $S_1$-$S_2$ dark plane, dependent on the orientation of the HWP.}
\label{fig:setup_channel}
\end{figure}

Our experiment is based on a realistic free-space channel of 1.6\,km length in the city of Erlangen [see Fig.~\ref{fig:setup_channel}(a)]. The channel is located in a truly urban environment. The optical beam path starting from the sender {\it (Alice)} runs past buildings, heated rooftops, heavy traffic, and forests. This inhomogeneous environment leads to significant turbulence, affecting phase and intensity. The beam is received by Bob, located in the 12th floor of a university building. In the following paragraph, we explain the details of our setup consisting of a robust source of polarization squeezed states of light at Alice and a detection stage at Bob.

At Alice, we generate CV polarization squeezed states in a  compact and stable setup by exploiting the Kerr nonlinearity of a single mode fiber~\cite{Heersink05, Milanovic07} [see Fig.~\ref{fig:setup_channel}(b)]. The source is able to operate without any need of controlled laboratory conditions. All components are built onto a breadboard with an area less than 0.3\,m$^2$. A commercial soliton laser (Origami, Onefive GmbH) emits femtosecond pulses ($200$\,fs) at a center wavelength of $\lambda_0=1559$\,nm and a repetition rate of $80$\,MHz.  Shorter wavelengths offer an enhanced collection efficiency after channel transmission at a given aperture size. Therefore, the sending and receiving optics of our free-space channel are optimized for wavelengths in the 800\,nm range, which also coincides with low atmospheric losses. Consequently, we use a periodically poled lithium-niobate crystal (MSHG1550-0.5-0.3, {\it Covesion Ltd.}) to convert the output of the laser to its second harmonic at $\lambda_1=780$\,nm and dichroic mirrors for subsequent spectral filtering. For the preparation of the CV polarization squeezed states we use both axes of a birefringent polarization maintaining photonic crystal fiber (NL-PM-700, zero dispersion wavelength $\lambda_{ZD}=700$\,nm, Crystal Fibre A/S). Because of the nonlinear Kerr effect in this fiber, we individually quadrature squeeze two orthogonally polarized pulses. These two pulses are overlapped to produce a circularly polarized state,  showing polarization squeezing. To achieve this, we have to precompensate the birefringence of the fiber with different optical path lengths in front of the fiber. We actively control the overlap and the phase between the two pulses by tapping 0.1\,\% of the signal after the fiber and controlling a piezoelectric actuator used in an interferometerlike configuration before the fiber.

For the description of the state of polarization the Stokes operators~\cite{Korolkova02} are used:
\begin{alignat*}{2}
\hat{S}_{0} &= \hat{a}^{\dag}_{H} \hat{a}_{H} + \hat{a}^{\dag}_{V} \hat{a}_{V} &&\mathrm{(total~intensity)},\\
\hat{S}_{1} &= \hat{a}^{\dag}_{H} \hat{a}_{H} - \hat{a}^{\dag}_{V} \hat{a}_{V} &&\mathrm{(horizontal~or~vertical~polarization)},\\
\hat{S}_{2} &= \hat{a}^{\dag}_{H} \hat{a}_{V} + \hat{a}^{\dag}_{V} \hat{a}_{H}  &&\mathrm{(}\pm45^{\circ}\mathrm{~polarization)},\mathrm{~and}\\
\hat{S}_{3} &= \imath(\hat{a}^{\dag}_{V} \hat{a}_{H} - \hat{a}^{\dag}_{H} \hat{a}_{V}) &&\mathrm{(left\hbox{-}~or~right\hbox{-}handed~polarization)},
\end{alignat*}
with the field operators for the horizontally and vertically polarized modes $\hat{a}_{H,V}$ and $\hat{a}^{\dag}_{H,V}$. In terms of these Stokes operators, our state has a mean value of 0 for  $\hat{S}_1$ and $\hat{S}_2$, while $\langle\hat{S}_3\rangle\gg0$. In Poincar{\'e} space, the plane perpendicular to $\hat{S}_3$ is called a dark plane in this case. The combined circularly polarized light pulses have a continuous wave equivalent optical power of $1.37$\,mW and show squeezing along one particular measurement direction within the $S_1$-$S_2$ dark plane $S_{\theta_\mathrm{sq}}$ and antisqueezing in the orthogonal direction $S_{\theta_\mathrm{sq}+90^\circ}$~\cite{Heersink05}. We measured the squeezing value  at Alice with an electrical spectrum analyzer at a sideband frequency of 12\,MHz, a resolution bandwidth of 300\,kHz, and a video bandwidth of 30\,Hz. For the measurements at Bob reported in this Letter, we measured a value of squeezing of $-2.4 \pm 0.1 $\,dB at Alice. The high antisqueezing value of $14.2 \pm 0.1$\,dB is a result of quantum noise and excess noise from photon-phonon interactions in the fiber. Finally, the optical beam is expanded by a telescope decreasing the divergence of the beam. The prepared states are distributed through the free-space channel.

At Bob, it is important to collect as much as possible of the incoming light field to minimize degrading loss. Simultaneously,  it is crucial to avoid any disturbance of the state of polarization. Therefore, we do not use a mirror but an achromatic lens with a diameter of 150\,mm and a focal length of $f=800$\,mm to collect the light. A half-wave plate (HWP) and a polarizing beam splitter (PBS) are placed slightly before the focal point, such that the beam can be directly focused onto two pin-diode (S3883, quantum efficiency $\eta=0.9$, {\it Hamamatsu Photonics K.K.}) detectors behind the PBS [see Fig.~\ref{fig:setup_channel}(c)]. The ac outputs of these custom made detectors are mixed with a 12\,MHz radio frequency local oscillator (arbitrary waveform generator 33250A, {\it Agilent Technologies Inc.}), amplified (wideband voltage amplifier DHPVA-100, {\it FEMTO Messtechnik GmbH}), low pass filtered (BLP-1.9, {\it Mini-Circuits Laboratory}, 3\,dB cutoff frequency 2.5\,MHz), and digitized by an analogue-to-digital converter (AD card, CompuScope CS1610, {\it GaGe a product brand of DynamicSignals LLC}) with a sampling rate of 10\,MSamples/s. The bandwidth of the dc outputs, that are sampled simultaneously with identical sampling rates is $150$\,kHz and thus sufficient for the atmospheric fluctuations. This setup allows us to measure the Stokes observables in the $S_1$-$S_2$ dark plane at a sideband frequency of 12\,MHz with a bandwidth of 2.5\,MHz (given by the used low pass filters) by calculating the difference of the two, still oversampled, ac signals. The measured Stokes observable within the $S_1$-$S_2$ dark plane can be chosen by rotating the HWP. The sum of the ac signals gives rise to the $S_0$ signal. The sum of the dc signals, from which we can calculate the respective transmission values, can be used as classical side information.
%

A turbulent atmosphere leads to beam wandering, distortion, and spreading and thus to continuous fluctuations of the intensity at the receiver. This directly results in a varying shot noise level during our measurements. It is important to account for this with a proper shot noise calibration and monitoring. We use the transmission value obtained as classical side information and sort all collected data according to the prevailing transmission value into bins with a width of $0.19$\,\% of the total optical power sent from Alice (corresponding to $2.6$\,$\mu$W). With this technique, we are able to compare the measured Stokes parameter variance to the corresponding shot noise variance.

\begin{figure}
\centering
\includegraphics[width=0.9\columnwidth]{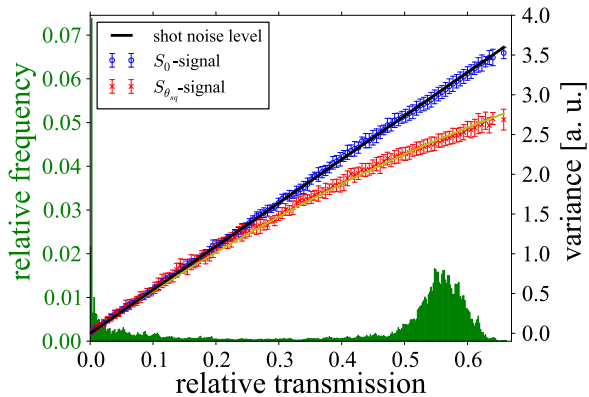}
\caption[]{Verification of transmitted squeezed states: Measurement of the polarization squeezed signal with additional artificially induced loss at the receiver. The  $S_0$ signal, which determines the shot noise limit, scales linearly with the transmission value (fitted in black). In good agreement to theory the polarization squeezed signal $S_{\theta_\mathrm{sq}}$ scales quadratically with the transmission value (fitted in yellow).  We show only every third data point for clarity. The green histogram shows the transmission statistics, which is a result of the natural fading channel and artificial attenuation.}
\label{fig:Abschw_sq}
\end{figure}

A reliable way to prove genuine detection of squeezed states is to compare the attenuation of the quantum states with its theoretical prediction. Squeezed states show a distinct behavior that can be discriminated from classical noise and rule out technical effects such as saturation of the detectors. We use both the natural fluctuation in transmission and additional attenuation of the beam imposed at the receiver to be able to measure the squeezed states at Bob for all transmission values down to 0 (see Fig.~\ref{fig:Abschw_sq}). The additional attenuation was achieved by moving an optically dense material into the beam at the receiver stage.
We carefully adjust the HWP to measure $S_{\theta_\mathrm{sq}}$ ($S_{\theta_\mathrm{sq}+90^\circ}$) by minimizing (maximizing) the measured variance. After the measurement we average the variances of the total intensity and the Stokes parameters over 10\,000 samples for each individual transmission bin and discard all bins containing fewer than 50\,000 samples. Thus, at least five variances and their mean value were obtained for each bin. To give a conservative estimate of the error of these measurements the error bar indicates 1 standard deviation of the deduced variances. The $S_0$ signal (blue circles) scales linearly with the transmission (black fit), which is in perfect agreement to a shot noise limited signal. The polarization squeezed signal $\hat S_{\theta_\mathrm{sq}}$ (red crosses) scales quadratically (yellow fit), as expected from theory~\cite{Bachor04}, and is below the shot noise level. The normalized green histogram visualizes the transmission statistics during this measurement.

The measurement of the Stokes observable $\hat S_{\theta_\mathrm{sq}+90^\circ}$ (see the Supplemental Material~\cite{sup})  shows antisqueezing as well as classical excess noise stemming mostly from photon-phonon interaction in the fiber used for squeezed light generation in the sender setup.

\begin{figure}
\centering
\includegraphics[width=0.9\columnwidth]{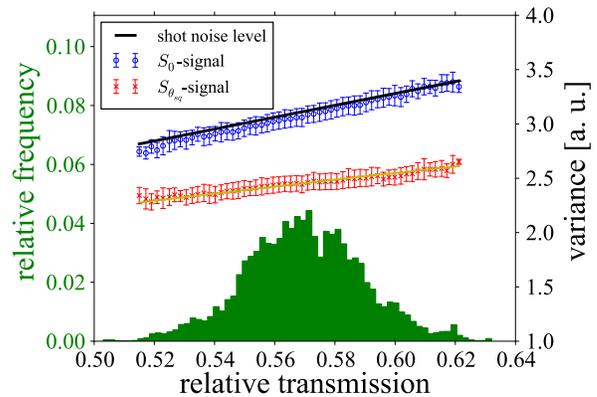}
\caption{Measurement of the polarization squeezed signal with natural transmission statistics of the free-space channel. The normalized green histogram shows the transmission statistics. With the state selection protocol, we can improve the squeezing value from  $-0.95$\,dB to $-1.08$\,dB relative to shot noise variance.}
\label{fig:sq}
\end{figure}

After verifying the genuine transmission of squeezed states, we investigated the natural transmission statistics of our free-space fading channel, which is mainly influenced by the clipping of the beam at the receiving aperture (the absorption in the atmosphere in clear weather conditions typically amounts to -0.1\,dB/km~\cite{Scarani09}, which contributes less than 4\% for a link of 1.6 km length). The beam spot size of the received beam is broadened due to the wave front distortions originating from the atmospheric influence. Additionally, the atmospheric turbulence leads to beam wandering. The transmission statistics is shown in Fig.~\ref{fig:sq} as a normalized green histogram. The shot noise variance (black line) is determined by the corresponding measurement shown in Fig.~\ref{fig:Abschw_sq}. The squeezing value for the squeezed state with the highest transmission value is $-1.08 \pm 0.04$\,dB below the shot noise variance, slightly increased compared to the average amount of squeezing of $-0.95 \pm 0.03$\,dB. Both values are obtained without any correction for losses. The enhancement by the postselection protocol which selects the transmission bin with the highest value depends on several parameters. This will be more pronounced if the investigated channel transmission  covers a broader range or if the highest possible channel transmission would be closer to the physical limit of 1. Finally, this postselection technique becomes most important if the initial squeezing is very high, as squeezing degrades in a highly nonlinear fashion with respect to losses.

The state selection protocol enables us to perform a quantum state reconstruction of a selected squeezed state after the noisy channel. Here, we can interpret the $S_1$-$S_2$ dark plane as the quadrature phase space as $\langle\hat S_3\rangle\gg0$.

\begin{figure}
\centering
\includegraphics[width=1\columnwidth]{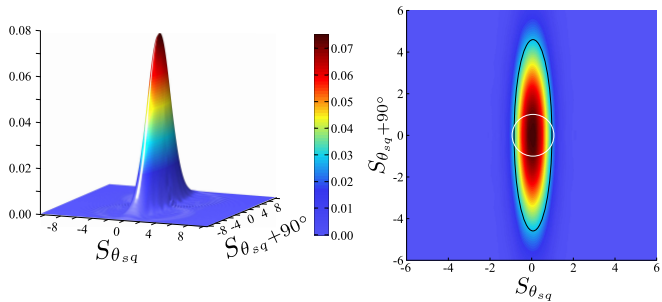}
\caption{The reconstructed Wigner function of a squeezed state after its transmission through the atmospheric free-space channel. On the left-hand side, we show the three-dimensional surface of the Wigner function. On the right-hand side, the same Wigner function is viewed from the top two dimensionally. The black and white lines visualize the $1/e$ contour for the squeezed state and the vacuum state, respectively.}
\label{fig:Wigner}
\end{figure}

We select the state with 55.2\% transmission. We chose this state because it offers a sufficiently good statistic for all measured angles. We measure the Stokes observables at 29 different angles between the squeezing direction $S_{\theta_\mathrm{sq}}$ and the antisqueezing direction $S_{\theta_\mathrm{sq}+90^\circ}$. At each angle, we use a total of $3.15\cdot10^5$ samples which are subsequently sorted into 251 bins to derive the corresponding tomograms. Accounting for the anticipated symmetry of the squeezed state, we mirror the acquired tomograms to attain a full tomographic dataset comprising 58 different measurement angles in the dark plane~\cite{Mueller12}. Based on these tomograms, we finally reconstruct the density matrix $\rho$ in the Fock state basis via a maximum likelihood algorithm~\cite{Vardi93,Hradil97,Lvovsky04}. Again, we do not correct the reconstructed state for losses. The reconstruction requires truncating the density matrix to an adequate maximum photon number, where in our case a matrix with dimension 54 x 54, i.e. with a maximum photon number of 53, proved to be fully reconstructible from 58 tomograms~\cite{Sych12}. Any matrix element of higher order only contributes to the state with a relative weight of less than $10^{-3}$ with respect to the largest matrix element, i.e. the vacuum state. From the reconstructed state we determine a purity $p=\mathrm{Tr}(\rho^2) =0.237$.
The essential sector of the reconstructed density matrix is shown in the Supplemental Material~\cite{sup}. A more intuitive grasp of the state's properties is provided by means of the Wigner function which we calculate~\cite{Qutip} from the reconstructed density matrix. The result is plotted in Fig.~\ref{fig:Wigner}.
In this representation, the comparison between the contour lines of the coherent vacuum state and the reconstructed state directly reveals the preservation of the squeezing.

We successfully distributed polarization squeezed states through a free-space channel. To the best of our knowledge we show the first distribution of continuous variable nonclassical states through the turbulent atmosphere. We presented a technique for taking into account a natural drifting shot noise variance. Furthermore, we enhanced the amount of squeezing of the state via a state selective protocol and reconstructed a selected state. This substantiates the feasibility of CV free-space quantum communication and paves the way for further free-space quantum communication experiments such as squeezing distillation~\cite{Heersink06-2}, CV entanglement distribution and distillation~\cite{Dong10}, or exploiting CV quantum states for atmospheric sensing.

The project was supported under FP7 FET Proactive by the integrated project Q-Essence and CHIST-ERA (Hipercom). The authors acknowledge support of the Erlangen Graduate School in Advanced Optical Technologies (SAOT) by the German Research Foundation (DFG) in the framework of the German Excellence Initiative.\\
The authors thank Ch. Wittmann, L. L. S\'anchez-Soto and D. Sych for fruitful discussions and our colleagues at the FAU computer science building for their kind support and for hosting the receiver station.\\
C. P. and B. H. contributed equally to this work.

\renewcommand{\thefigure}{\Alph{figure}}
\setcounter{figure}{0}
\cleardoublepage
\section{Supplemental Material}
\subsection{Measurement of the antisqueezed Stokes observable}
\begin{figure}[h]
\centering
\includegraphics[width=0.9\columnwidth]{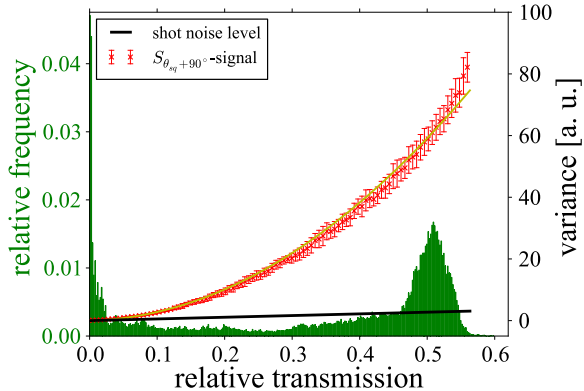}
\caption[]{Measurement of the polarization antisqueezed signal with classical excess noise and additional artificially induced loss at the receiver. Note that the classical excess noise is very high due to photon-phonon interaction in the fiber used at Alice. The signal scales quadratically (fitted in yellow) as expected from theory. We draw only every third data point for clarity. The green histogram shows the transmission statistics for this measurement.}
\label{fig:Abschw_antisq}
\end{figure}

To verify the proper transmission of squeezed states we show a measurement of the squeezed observable $S_{\theta_\mathrm{sq}}$ over a wide transmission range in the main letter (see Fig.~2 therein). Additionally, we measured the Stokes observable $\hat S_{\theta_\mathrm{sq}+90^\circ}$ and show the result in Fig.~\ref{fig:Abschw_antisq}.
As mentioned in the main Letter, this observable exhibts antisqueezing, but is dominated by classical excess noise from photon-phonon interaction in the fiber that we use for squeezed light generation. The variance of $\hat S_{\theta_\mathrm{sq}+90^\circ}$ (red crosses) scales quadratically with the transmission value (fitted in yellow), while the black line indicates the shot noise limit as given by the fitted curve of Fig.~2 of the main letter.
\\
\subsection{Discussion of the reconstructed density matrix}
\begin{figure}
\centering
\includegraphics[width=0.9\columnwidth]{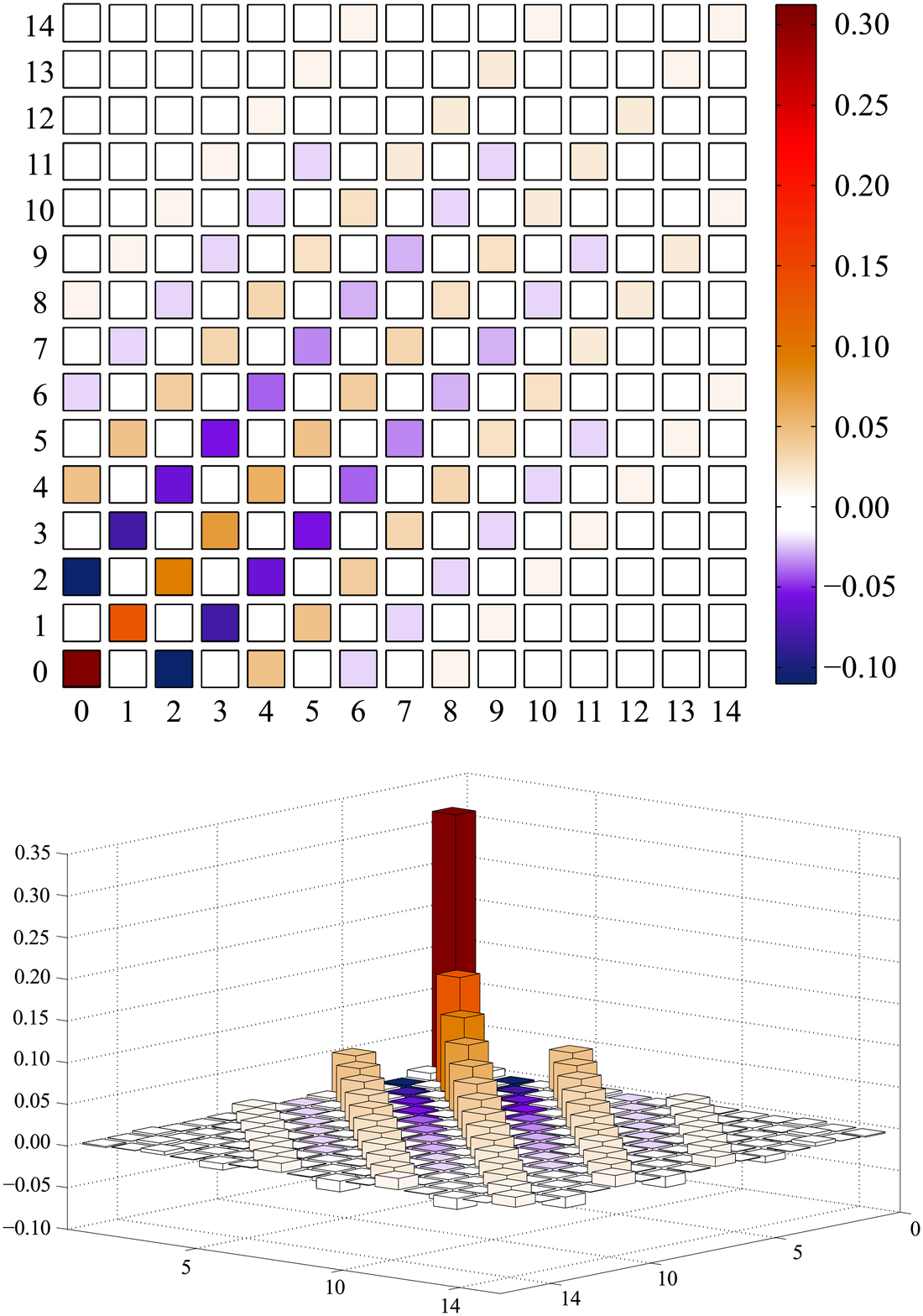}
\caption[]{Top view and bar chart of the reconstructed density matrix. The integers on the axes refer to photon number states.}
\label{fig:rho}
\end{figure}

In the main Letter we discuss the reconstruction of the density matrix of a transmitted squeezed state and show the Wigner function, which is calculated from that density matrix in Fig.~4. Here we directly show the essential part of the density matrix (see Fig.~\ref{fig:rho}). The density matrix is dominated by the contributions on the diagonal, but an important signature of squeezing can be seen in the non-zero off-diagonal matrix elements. If only matrix elements of the diagonal would contribute to the density matrix, the resulting Wigner function would be radially symmetric.

During the squeezing process correlated pairs of photons are generated in the sidebands aside of the signal carrier. These correlations manifest themselves in contributions at matrix elements with both even indices (e.g. $\rho_{0,2},~\rho_{4,2}$). In the reconstructed density matrix however, also matrix elements with both odd indices (e.g. $\rho_{1,3},~\rho_{5,1}$) are contributing to the state. This is due to (classical) excess noise.
\end{document}